\documentclass{elsart}
\usepackage{amssymb}
\usepackage{epsfig}
\usepackage{graphicx,color}
\journal{Int. J. for Numerical and Analytical Methods in Geomechanics}
\date{01/30/2009}

\begin{document}

\title{On the capillary stress tensor in wet granular materials.}

\author[label1]{L. Scholt\`es},
\author[label2]{P.-Y.~Hicher},
\author[label3]{F. Nicot},
\author[label1]{B. Chareyre\corauthref{cor1}},
\corauth[cor1]{Corresponding author}
\ead{bruno.chareyre@hmg.inpg.fr}
\author[label1]{F. Darve}.

\address[label1]{Laboratoire 3S-R (Sols, Solides, Structures - Risques), Grenoble Universit\'es, Domaine Universitaire BP 53, 38041 Grenoble cedex 9, France.}
\address[label2]{Institut de Recherche en G\'enie Civil et M\'ecanique GeM CNRS, Ecole Centrale de Nantes, BP 92101, 44321 Nantes Cedex 3, France.}
\address[label3]{Cemagref - Unit\'e de recherche Erosion Torrentielle Neige et Avalanches, Domaine Universitaire BP 76, 38402 Saint-Martin d'H\`eres cedex, France.}



\begin{frontmatter}

\begin{abstract}
This paper presents a micromechanical study of unsaturated granular media in the pendular regime, based upon numerical experiments using the discrete element method, compared to a microstructural elastoplastic model. Water effects are taken into account by adding capillary menisci at contacts and their consequences in terms of force and water volume are studied. Simulations of triaxial compression tests are used to investigate both macro and micro-effects of a partial saturation. The results provided by the two methods appear to be in good agreement, reproducing  the major trends of a partially saturated granular assembly, such as the increase in the shear strength and the hardening with suction. Moreover, a capillary stress tensor is exhibited  from  capillary  forces by using homogenisation techniques. Both macroscopic and microscopic considerations emphasize an induced anisotropy of the capillary stress tensor in relation with the pore fluid distribution inside the material. In so far as the tensorial nature of this fluid fabric implies shear effects on the solid phase associated with suction, a comparison has been made with the standard equivalent pore pressure assumption. It is shown that water effects induce microstrural phenomena that cannot be considered at the macro level, particularly when dealing with material history. Thus, the study points out that unsaturated soil stress definitions should include, besides the macroscopic stresses such as the total stress, the microscopic interparticle stresses such as the ones resulting from capillary forces, in order to interpret more precisely the implications of the pore fluid on the mechanical behaviour of granular materials. [published, DOI: 10.1002/nag.767]
\end{abstract}

\end{frontmatter}
\section{INTRODUCTION}

Macroscopic properties of granular materials such as soils depend on particle interactions. In dry granular materials, interparticle forces are related to the applied external loads as different studies have shown \cite{NicotDarve2005, Cundall1979}. In unsaturated soils subjected to capillary effects, new features must be accounted for in order to understand properly their behaviour. The presence of water leads to the formation of water menisci between neighboring grains, introducing new interparticle forces. The effects of these forces depend on the degree of saturation of the medium. For low water content level corresponding to disconnected liquid bridges between grains, capillary theory allows the force induced by those bridges to be linked to the local geometry of the grains and to the matric suction or capillary pressure inside the medium \cite{Fisher1926}. Since the disconnected menisci assumption is not valid for high water content levels due to water percolation, we consider here only the unsaturated state where the discontinuity of the water phase can be assumed, the so-called pendular regime.\\

There has been a wide debate on the various interpretations for the mechanical behaviour of unsaturated soils. At early stages of soil mechanics, Terzaghi \cite{Terzaghi1925} first introduced the concept of effective stress for the particular case of saturated soils enabling the conversion of a multiphase porous medium into a mechanically equivalent single-phase continuum. In unsaturated soils, water induced stresses are still debated. The common practice \cite{Bishop1959, FredlundMorgenstern1978} is to use the suction or a modified version as a second stress variable within a complete hydro-mechanical framework. An alternative method is introduced to develop homogenisation techniques in order to derive stress-strain relationships from forces and displacements at the particle level as proposed in \cite{NicotDarve2005} for dry granular materials. The basic idea is to consider the material as represented by a set of micro-systems, postulating that the behaviour of a material volume element depends on the intergranular interactions belonging to this volume. We propose here to extend this micro-mechanical approach to unsaturated granular materials as proposed by Li \cite{Li2003}, Jiang et al. \cite{Jiang2004} or Lu and Likos \cite{LuLikos2006}.\\

Along these lines we present two micromechanical models which take into account capillary forces. The first one is a three dimensional numerical model based on the Discrete Element Method (hereafter designed as the DEM model) pioneered by Cundall and Strack \cite{Cundall1979}, and the second one is an analytical model (hereafter designed as the microstructural model) recently proposed by Hicher and Chang \cite{HicherChang2006}.

The microstructural model is a stress-strain relation which considers interparticle forces and displacements. Thanks to analytical homogenisation/localisation techniques, the macroscopic relation between stress and strain can be derived.

In the DEM model, a granular medium is modelled by a set of grains interacting according to elementary laws. Direct simulations are carried out on grain assemblies, computing the response of the material along a given loading path.

By studying their effects under triaxial loading, we investigate capillary forces implications at the macroscopic level, and offer an insight into the unsaturated soil stress framework by introducing a capillary stress tensor as a result of homogenization techniques.

\section{UNSATURATED SOIL STRESSES}
\subsection{Macroscopic views}
Macroscopic interpretations of the mechanical behaviour of unsaturated soils have been mainly developed in the framework of elastoplasticity \cite{NuthLaloui2007}. Most of these models consider that the strain tensor is governed by the net stress tensor $\sigma_{ij}-u_a\delta_{ij}$ ($u_a$ being the pore air pressure) and the matric suction or capillary pressure $u_a-u_w$ ($u_w$ being the pore water pressure) inside the medium \cite{AlonsoGens1990, WheelerSivakumar1995}. In particular, they consider a new yield surface, called Loading Collapse (LC) surface in the plane (($\sigma_{ij}-u_a\delta_{ij}$),($u_a-u_w$)) which controls the volume changes due to the evolution of the degree of saturation for a given loading path. As a matter of fact, all these formulations can be considered as extensions of the relationship initially proposed by Bishop and Blight \cite{Bishop1963} for unsaturated soils:
\begin{equation}
\sigma'_{ij} = (\sigma_{ij} - u_a\delta_{ij})+\chi(S_r)(u_a-u_w)\delta_{ij}
\end{equation}
where $\chi(S_r)$ is called the effective stress parameter or Bishop's parameter, and is a function of the degree of saturation $S_r$ of the medium ($\chi=0$ for a dry material, $\chi=1$ for a fully saturated material).\\

Obviously, since the effective stress principle is by definition a macroscopic concept, several authors (Lu and Likos \cite{LuLikos2006} or Li \cite{Li2003}) have proposed to use a micromechanical approach for the effective stress principle. In order to further study this micromechanical approach to study unsaturated soil stresses, we propose here a micromechanical analysis of the problem, examining the local water induced effects through a set of simulated laboratory experiments.

\subsection{Micromechanical interpretation}
Let us consider a Representative Volume Element (RVE) of a wet granular material, subjected to an assigned external loading. When the water content decreases inside a saturated granular sample, the air breaks through at a given state. The capillary pressure ($u_a-u_w$) corresponding to that point is called the air-entry pressure, and strongly depends on the pore sizes. Thereafter, the sample becomes unsaturated and capillary forces start to grow due to interface actions between air and water. Since the the gaseous phase is discontinue, this is the capillary regime. From this state, a constant decrease in the degree of saturation corresponds to a gentle increase in pore water pressure. The pendular regime starts when the water phase is no longer continuous. In this state, fluids equilibrium is obtained by the vapor pressure. Analytical and experimental results \cite{Haines1925, Fisher1926} demonstrate that capillary effects at particle contacts produces a kind of bond between particles as a result of menisci (Fig.\ref{fig1}). Liquid bridges may form between some pairs of adjoining particles not necessarily in contact, generating an attractive capillary force between the bonded particles. If the drying process continues, these water bridges begin to fail, starting from the non-contacting grains, until the complete disappearance of capillary forces inside the assembly.\\

As the pendular regime is considered throughout this paper, water is considered to be solely composed of capillary menisci: each liquid bridge is assumed to connect only two particles. Therefore, two types of forces coexist within the granular medium. For dry contacts, a contact force develops between contacting granules. This repulsive force, that is a function of the relative motion between the contacting grains, is usually well described by an elastoplastic contact model. For water bonded particles, a specific attractive force exists. This water-induced attractive interaction can be described by a resulting capillary force, rather than by a stress distribution as mentionned by Haines \cite{Haines1925} or Fisher \cite{Fisher1926}. This capillary force is a function of the bridge volume, of the size of particles, and of the fluid nature (see section 3.1.1 for the details). The objective of this section is to derive, in a simple manner, an expression relating the overall stress tensor within the RVE to this internal force distribution.\\

For this purpose, the Love \cite{Love1927} static homogenisation relation is used. This relation expresses the mean stress tensor $\sigma$ within a granular volume $V$ as a function of the external forces $\vec{F}^{ext,p}$ applied to the particles $p$ belonging to the boundary $\partial V$ of the volume:
\begin{equation}
\sigma_{ij} = \frac{1}{V} \sum_{p \epsilon \partial V} F_i^{ext,p} x_j^p
\end{equation}
where $x_j^p$ are the coordinates of the particle $p$ with respect to a suitable frame. It is worth noting that this relation is valid whatever the nature of the interactions between grains.\\

Taking into account the mechanical balance of each particle of the volume $V$ (including the boundary $\partial V$), Eq.(2) can be written as:
\begin{equation}
\sigma_{ij} = \frac{1}{V} \sum_{p=1}^N \sum_{q=1}^N F_i^{q,p} l_j^{q,p}
\end{equation}
where $N$ is the number of particles within the volume, $\vec{F}^{q,p}$ is the interaction force exerted by the particle $q$ onto the particle $p$, and $\vec{l}^{q,p}$ is the branch vector pointing from particle $q$ to particle $p$ ($\vec{l}^{q,p} = \vec{x}^p - \vec{x}^q$).\\

As we consider partially saturated granular media, two independent kinds of interparticle forces can be distinguished:
\begin{enumerate}
\item[(i)] if particles $p$ and $q$ are in contact, a contact force $\vec{F}_{cont}^{q,p}$ exists.
\item[(ii)] if particles $p$ and $q$ are bonded by a liquid bridge, a capillary force $\vec{F}_{cap}^{q,p}$ exists.
\end{enumerate}

Actually, depending on the local geometry, a liquid bond can exist between two grains in contact. In that case, solid contacts are surrounded by the continuous liquid phase providing the simultaneity of contact and capillary forces. The two contributions have therefore to be accounted for by summation.\\

Finally, in all cases and for any couple $(p,q) \epsilon [1,N]^2$, it can be written that:
\begin{equation}
\vec{F}^{q,p} = \vec{F}_{cont}^{q,p} + \vec{F}_{cap}^{q,p}
\end{equation}
Thus, by combining Eqs.(2) and (4), it follows that:
\begin{equation}
\sigma_{ij} = \frac{1}{V} \sum_{p=1}^N \sum_{q=1}^N F_{cont,i}^{q,p} l_j^{q,p} + \frac{1}{V} \sum_{p=1}^N \sum_{q=1}^N F_{cap,i}^{q,p} l_j^{q,p}
\end{equation}

As a consequence, Eq.(5) indicates that the stress tensor is split into two components:
\begin{equation}
\sigma_{ij} = \sigma_{ij}^{cont} + \sigma_{ij}^{cap}
\end{equation}

\begin{enumerate}
\item[-] A first component $\sigma_{ij}^{cont} = \frac{1}{V} \sum_{p=1}^N \sum_{q=1}^N F_{cont,i}^{q,p} l_j^{q,p}$ accounting for the contact forces transmitted along the contact network.
\item[-] A second component $\sigma_{ij}^{cap} = \frac{1}{V} \sum_{p=1}^N \sum_{q=1}^N F_{cap,i}^{q,p} l_j^{q,p}$ representing the capillary forces existing within the assembly.
\end{enumerate}

It is to be noted that $\sigma_{ij}^{cont}$ is a stress quantity standing for intergranular contact forces in the same way as in saturated or dry conditions. Considering the concept as initially introduced by Terzaghi, $\sigma_{ij}^{cont}$ plays the role of the so-called effective stress by governing soil deformation and failure. Besides, $\sigma_{ij}^{cap}$ is the tensorial attribute to capillary water effects or suction, by extension. By analogy with Eq.(1), we can therefore define a microstructural effective stress
\begin{equation}
\sigma_{ij}^{cont} = \sigma_{ij} - \sigma_{ij}^{cap}
\end{equation}
where $\sigma_{ij}$ could be affiliated to net stress, representing the apparent stress state in the material. Compared with Eq.(1), where the effect of water is intrinsically isotropic, $\sigma_{ij}^{cap}$ implies a tensorial attribute to the water effects.\\

In fact, in both terms $\sigma_{ij}^{cont}$ and $\sigma_{ij}^{cap}$, a fabric tensor can emerge from the summation \cite{Love1927, Christofferson1981, Rothenburg1981}. The fabric tensor is useful to characterize the contact anisotropy of the assembly, which is known as a basic feature of granular assemblies. In dry granular materials, an induced anisotropy can develop when a deviatoric stress loading is applied. In partially saturated assembly, due to the possibility of interactions without contact, the conclusion is not so trivial. As an illustration, if we restrict our analysis to spherical particles \cite{NicotDarve2005}, it can be inferred that:
\begin{equation}
\sigma_{ij}^{cap} = \frac{1}{V} \sum_{p=1}^N \sum_{q=1}^N F_{cap}^{q,p} l^{q,p} n_i^{p,q} n_j^{p,q}
\end{equation}
This relation points out that, contrary to the contact term where the fabric tensor $\frac{1}{N} \sum_{q=1}^N n_i^{p,q} n_j^{p,q}$ is directly linked to the induced anisotropy (a contact is associated with a force), two causes can be invoked for the capillary term. First, the distribution of the liquid bonds can be anisotropic. Secondly, the geometry of the bonds being obviously dependent on the local geometry, it is possible that the distribution of both terms $F_{cap}^{q,p}$ and $l^{q,p}$ is also anisotropic. This is significant because the anisotropic attribute yields shear effects associated with pore fluid, which could mostly influence material behaviour.\\

In order to enrich our discussion, we present numerical investigations of these features in the following sections using both DEM and micromechanical simulations.

\section{MICROSTRUCTURAL INVESTIGATION ON THE CAPILLARY STRESS TENSOR USING DEM}
We present here a numerical analysis of the stress variables using a micromechanical model based upon the Discrete Element Method initially introduced by Cundall and Strack \cite{Cundall1979}. This technique starts with basic constitutive laws between interacting particles and can provide a macroscopic response of a particle assembly due to loading changes at the boundaries. Each particle of the material is a rigid sphere identified by its own mass, $m$, radius, $R$ and moment of inertia, $I_0$. For every time step of the computation, interaction forces between particles, and consequently resulting forces acting on each of them, are deduced from sphere positions through the interaction law. Newton's second law is then integrated through an explicit second order finite difference scheme to compute the new sphere positions.

\subsection{The discrete element model}
A 3D software called YADE (Yet Another Dynamic Engine), Kozicki and Donz\'e \cite{YADE}, has been enhanced in order to properly simulate partially saturated granular material features.

\subsubsection{Inter-particle behaviour}

The contact interaction is described by an elastic-plastic relation between the force $F$ and the relative displacement $U$ of two interacting particles. A normal stiffness $K_n$ is therefore defined to relate the normal force $F_n$ to the intergranular normal relative displacement $U_n$ :
\begin{equation}
F_n = \left\{ \begin{array}{ll} K_n U_n \quad \textrm{if} \quad U_n\leq0\\
				0 \quad \textrm{if} \quad U_n > 0
	      \end{array} \right.
\end{equation}
and a tangential stiffness $K_t$ allows us to deduce the shear force $F_t$ induced by the incremental tangential relative displacement $dU_t$; this tangential behaviour obeying the Coulomb friction law:
\begin{equation}
\left\{ \begin{array}{ll} dF_t = -K_t dU_t\\
			  F_t^{max} = -\mu F_n
	\end{array} \right.
\end{equation}
where $\mu$ is the Coulomb friction coefficient defined as $\mu = tan(\phi)$, with $\phi$ the intergranular friction angle.\\

In the work presented here, $K_n$ and $K_t$ are dependent functions of the interacting particle sizes and of a characteristic modulus of the material denoted as $E$:
\begin{equation}
\left\{ \begin{array}{ll} K_n = 2E \frac{R1R2}{(R1+R2)}\\ 
			  K_t = \alpha\,K_n
	\end{array} \right.
\end{equation}
This definition results in a constant ratio between $E$ and the effective bulk modulus of the packing, whatever the size of the particles.\\

For simplicity, we assume that the water inside the sample is solely composed of capillary water as defined in the pendular state, with a discontinuous liquid phase.

\begin{figure}
\centering
\includegraphics[width=130mm]{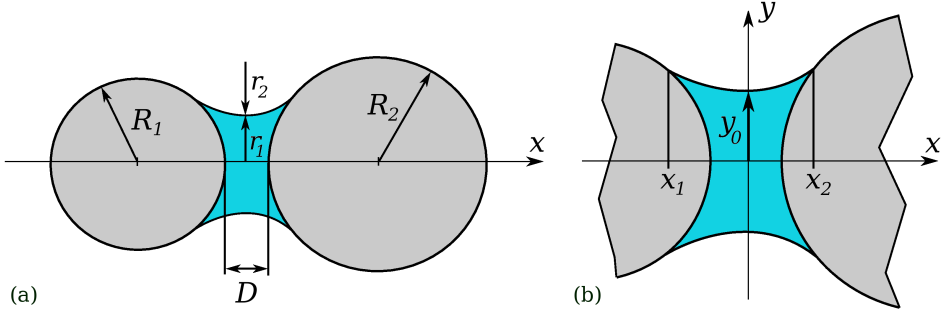}
\caption{Illustration of a liquid bridge between two particles of unequal sizes: (a) global geometry, (b) details of the bridge.}
\label{fig1}
\end{figure}
Much attention has been given to these pendular liquid bridges, \cite{Hotta1974, Lian1993, Willet2000}. Their exact shape between spherical bodies is defined by the Laplace equation, relating the pressure difference $\Delta u = u_a - u_w$  across the liquid-gas interface to the mean curvature of the bridge and the surface tension of the liquid phase $\gamma$:
\begin{equation}
\Delta u = \gamma \left( \frac{1}{r_1} + \frac{1}{r_2} \right)
\end{equation}

In the Cartesian coordinates of Fig.\ref{fig1}(b) the two curvature radii $r_1$ and $r_2$ (Fig.\ref{fig1}(a)), are given by:
\begin{equation}
\frac{1}{r_1} = \frac{1}{y(x)\sqrt{1+y'^2(x)}}
\end{equation}
and
\begin{equation}
\frac{1}{r_2} = \frac{y''(x)}{(1+y'^2(x))^{3/2}}
\end{equation}
where $y(x)$ defines the profile of the liquid-gas interface curve. The $x$ axis coincides with the axis of symmetry of the bridge, passing through the centers of the connected spheres (Fig.\ref{fig1}(b)).
According to the Laplace equation, the profile of the liquid bridge is thus related to the capillary pressure $\Delta u$ through the following differential equation:
\begin{equation}
\frac{\Delta u}{\gamma}(1+y'^2(x))^{3/2} + \frac{1+y'^2(x)}{y(x)} - y''(x) = 0
\end{equation}
The corresponding liquid bridge volume $V$ and intergranular distance $D$ can be obtained by considering the $x$-coordinates ($x_1$ and $x_2$) of the three-phases contact lines defining the solid-liquid-gas interface, as defined by Souli\'e et al.,\cite{Soulie2006} :
\begin{equation}
\begin{array}{cc}
V = \pi \int_{x_1}^{x_2} y^2(x)dx - \frac{1}{3} \pi R_1^3 (1-acos(x_1))^2(2+acos(x_1))\\
 - \frac{1}{3} \pi R_2^3 (1-acos(x_2))^2(2+acos(x_2))
\end{array}
\end{equation}
and
\begin{equation}
D = R_2(1-acos(x_2))+x_2+R_1(1-acos(x_1))-x_1
\end{equation}

The capillary force due to the liquid bridge can be calculated at the profile apex $y_0$ according to the `gorge method' \cite{Hotta1974} and consists of a contribution of the capillary pressure $\Delta u$ as well as the surface tension $\gamma$:
\begin{equation}
F_{cap} = 2\pi y_0\gamma + \pi y_0^2\Delta u
\end{equation}

\begin{figure}
\centering
\includegraphics[width=75mm]{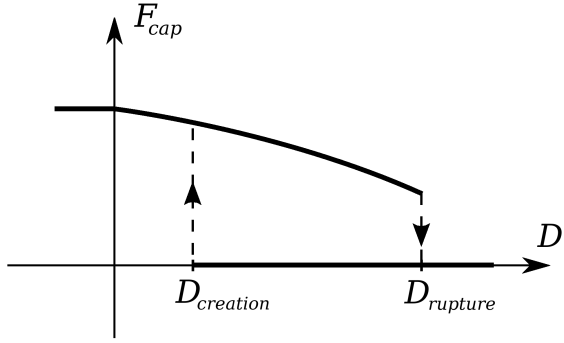}
\caption{Evolution of the capillary force $F_{cap}$ with the intergranular distance $D$ for a given suction value: a meniscus can form for $D < D_{creation}$ and breaks off for $D > D_{rupture}$.}
\label{fig2}
\end{figure}

The relation between the capillary pressure and the configuration of the capillary doublet is thus described by a system of non-linear coupled equations (15, 16, 17, 18) where the local geometry ($D$) and water volume arise as a result of the solved system \cite{Soulie2006}. So, to account for capillarity in the code, an interpolation scheme on a set of discrete solutions of the Laplace equation has been developed in order to link directly the capillary pressure to the capillary force and water volume of the liquid bridge for a given grain-pair configuration. This results in a suction-controlled model where, at every time-step during the simulation, capillary forces and water volumes ($F_{cap}, V$)  are computed based upon the microstructure geometry $D$ and the imposed suction level $\Delta u$.
\begin{equation}
( F_{cap}, V ) = \Im (\Delta u; D)
\end{equation}
A schematic diagram of the implemented capillary law is shown in Fig.\ref{fig2} for a given value of the suction.
In this paper, the choice was made to define the appearance of a bridge when particles come strictly in contact ($D_{creation} =0$), neglecting the possibility of adsorbed water effects. The capillary force is considered constant for the range of the elastic deformation ($D \leq 0$), assuming local displacements to be very small compared to particle radii. Let us note that the formulation intrinsically defines the distance from which the meniscus breaks off as depending on the given capillary pressure and on the local geometry. This maximum distance $D_{rupture}$ corresponds to the minimum $D$ value from which the Laplace equation has no solution.

\subsubsection{Stress tensors}

Since this study covers the macroscopic and microscopic aspects of unsaturated granular media, stress tensors are calculated by both macro and micro-methods. The macro-method is the conventional way used in laboratory to measure stresses in experiments, that is to say: 
\begin{equation}
\sigma_{ij} = (\sum F_i)/S_j
\end{equation}
where $F_i$ is the normal force acting on the boundary, and $S_j$ the surface of the boundary oriented by the normal direction $j$. This stress is equivalent to net stress $(\sigma_{ij} - u_a\delta_{ij})$ used in unsaturated soils mechanics, with $u_a$ used as the reference pressure because the pore air pressure is effectively zero in many practical problems (as well as in this study).\\

As seen in section 2.2, two other stress tensors can be considered through homogenization techniques: the intergranular stress tensor $\sigma^{cont}$ computed from intergranular forces, and the capillary stress tensor $\sigma_{ij}^{cap}$ computed from capillary forces.


\subsubsection{Sample description and testing programme}

\begin{figure}
\centering
\includegraphics[width=120mm]{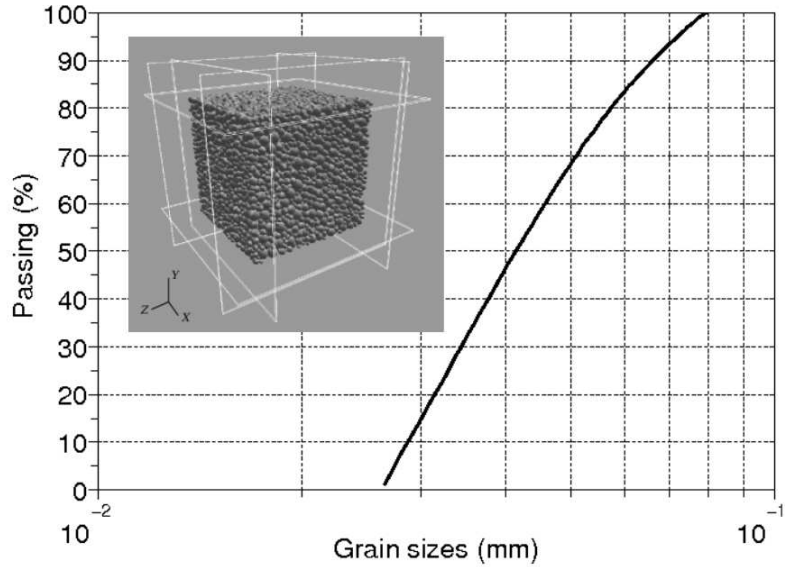}
\caption{Sample description.}
\label{fig3}
\end{figure}

The studied particle assembly is a 1 mm length cubic sample composed of 10000 spheres, with a grain size distribution ranging from 0.025 mm to 0.08 mm, as shown in Fig.\ref{fig3}, and a porosity of 0.385. The input parameters are listed in table 1, referring to equation (10).
\begin{table}[h]
\begin{center}
\begin{tabular}[c]{c c c c}
\hline
Global Modulus	& $\frac{K_t}{K_n}$	& Friction angle\\
$E$ (MPa)	& $\alpha$		& $\phi$ (deg.)  \\
\hline
150		& 0.5			& 30 \\
\hline
\label{Table1}
\end{tabular}
\caption{DEM model parameters}
\end{center}
\end{table}

The sample was prepared by an isotropic-compaction technique, which can be described in two stages.
\begin{itemize}
\item[(a)] All particles are randomly positioned inside a cube made up of 6 rigid walls in such a manner that no overlap/contact force develops between any pair of particles. The interparticle friction angle is set to a small value (the smaller the friction angle, the denser the assembly. A value of 0.2 degree has been chosen here) and particle radii are then homogeneously increased, whereas boundary walls stay fixed. The process is run until the confining pressure (15 kPa) is reached and equilibrium between the internal stress state and the external load is satisfied \cite{Mahboubi1996}.
\item[(b)] The interparticle friction coefficient is then changed to a value classically used in DEM simulation to reproduce an acceptable shear strength (30 degree) and boundary walls are servo-controlled in displacement to keep the equilibrium state.
\end{itemize}

Starting from the initially stable configuration, a given suction is then applied. In this case, capillary forces (Eq.(18)) are added to all existing contact forces ($D_{creation} = 0$, see figure \ref{fig2}). This process simulates the appearance of liquid bridges at body contacts as it would take place during a capillary condensation when the relative humidity of the surrounding air is increased, and coincides with a wetting of the material. The sample then reaches a new equilibium state from which different stress paths can be imposed. Note that capillary forces are assumed to be zero between a sphere and a wall.\\

Suction-controlled triaxial compression tests have been carried out on the generated specimen which was taken up from the initial state of $15\,kPa$ to upper confining pressures by isotropic compaction through wall displacements. The capillary pressure value ($10\,kPa$) has been chosen so as liquid bridge volumes are small enough to avoid the possibility of interconnected liquid bridges, and hence to ensure the pendular regime in the medium. The water content can be simply computed as the sum of all the liquid bridge volumes. In this case, the associated initial degree of saturation of the sample is about $20\,\%$. A constant compression rate is then applied in the axial direction, controlling the lateral walls in displacement to keep the confining pressure constant. To keep the quasi-static assumption, the loading rate was fixed sufficiently small so as the normalized mean resultant force on particles (which is 0 at static equilibrium) does not exceed $1\,\%$ at each loading step.

\subsection{Stress tensor analysis}
\subsubsection{Macroscopic evidence for the capillary stress tensor}
\begin{figure}
\centering
\includegraphics[width=100mm]{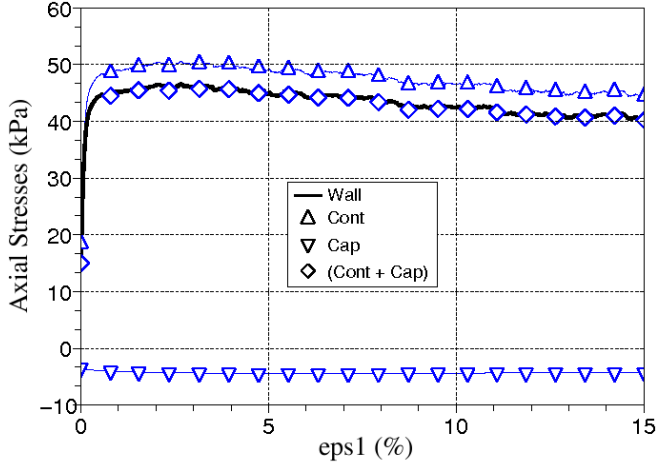}
\caption{Axial stresses resulting from elastic forces ($\sigma_{cont}$) and capillary forces ($\sigma_{cap}$), as well as the external load on the wall ($\sigma_{wall}$) under a suction-controlled triaxial compression.}
	\label{fig4}
\end{figure}

The quasi-static assumption ensures the equilibrium of the internal stress state and external load to be satisfied over every loading path. 
The internal stress has therefore to be the sum of the repulsive stress due to the elastic forces $F_n$ and the tensile one due to the capillary forces $F_{cap}$, as presented by Eq.(6) in subsection 2.2.
\begin{equation}
\sigma_{ij} = \sigma_{ij}^{cont} + \sigma_{ij}^{cap}
\end{equation}
As shown in figure \ref{fig4} for a triaxial loading path under a given confining pressure of 15kPa, this additivity is perfectly verified.
The quasi-static assumption is numerically confirmed and the existence of a capillary stress tensor is therefore proved for the case of wet granular materials.

\subsubsection{Capillary stress tensor analysis}

This suction induced stress leads to some questions about its structure and more generally on water distribution inside the material. In fact, classical considerations of unsaturated materials often assimilate suction effects to an equivalent pressure which consequently acts in the medium independently from its anisotropy (the "hydraulic" component of the effective stress is generally considered as an isotropic tensor).\\

\begin{figure}
	\centering
	\includegraphics[width=100mm]{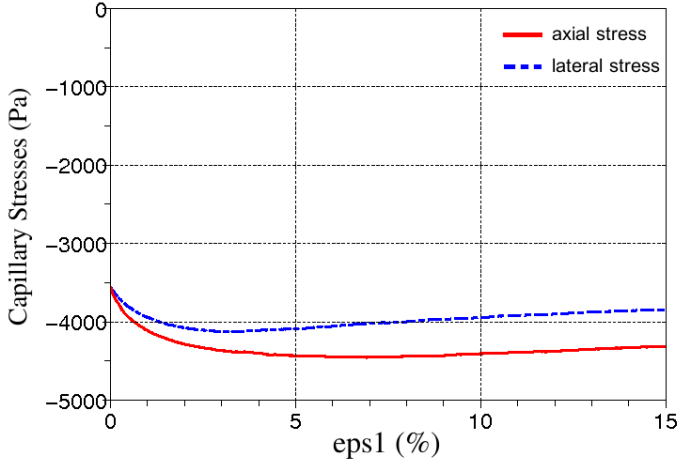}
 	\caption{Evolution of the principal capillary stress tensor components during a suction-controlled triaxial compression test ($P_0 = 15 kPa$, $u = 10 kPa$).}
	\label{fig5}
\end{figure}

Computing the principal components of this capillary stress tensor along the deviatoric loading path of a triaxial test compromises this assumption as shown in Fig.\ref{fig5}.
This is all the more remarkable in that the model ensures a uniform distribution of the capillary pressure inside the medium. It is clear that for the initial state corresponding to an isotropic configuration of the assembly, the capillary stress tensor is almost spherical with an initial mean value of about 3.6 kPa for both axial and lateral components. Nevertheless, the anisotropy rapidly evolves with the one induced by loading, providing a difference between the principal components. If we define $\alpha = 2 \frac{\sigma_1^{cap} - \sigma_2^{cap}}{(\sigma_1^{cap}+\sigma_2^{cap})}$ as a representative index of the tensor sphericity: $\alpha = 0$ for the initial isotropic state, and then slightly evolves to a quasi-constant value of $0.12$ from the $7\,\%$ deformation level until the final $15\,\%$ computed state.\\

\begin{figure}
\centering
\includegraphics[width=140mm]{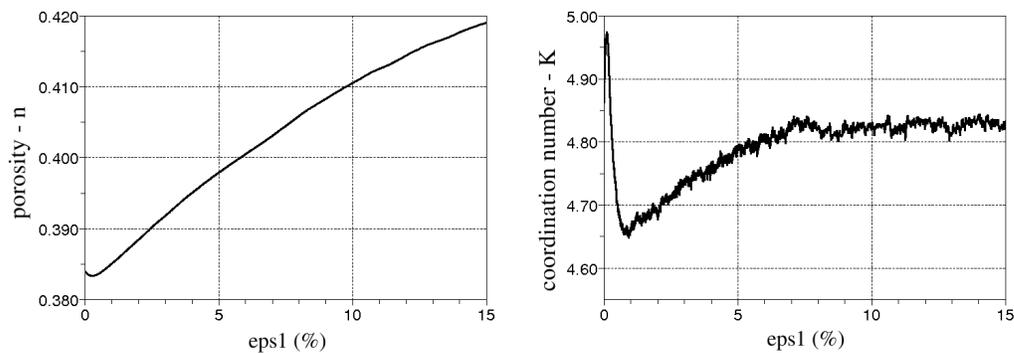}
\caption{Evolutions of the porosity and average number of contacts by particles $K$ under a suction-controlled triaxial compression ($P_0 = 15\,kPa$, $u = 10\,kPa$).}
	\label{fig6}
\end{figure}

The causes of this evolution can be analysed by examining the volumetric deformation of the sample (Fig.\ref{fig6}(a)) and the associated packing rearrangement through the average number of contacts by particle, $K$, (Fig.\ref{fig6}(b)).
As these global considerations have to be completed by a micromechanical analysis in order to gain a clear insight into the microstructural origins of the phenomenon, Fig.\ref{fig6} will be commented in the following section.

\subsubsection{Micromechanical investigation}

Here we develop a micromechanical analysis of the stress variables by considering both contact and liquid bridge distributions through the whole assembly. As the local interactions (dry contacts or menisci interactions) involve normal directions, a database can be defined in terms of orientations for all the grains of the sample. It is proposed  to examine this normal direction network.\\

\begin{figure}
\centering
\includegraphics[width=70mm]{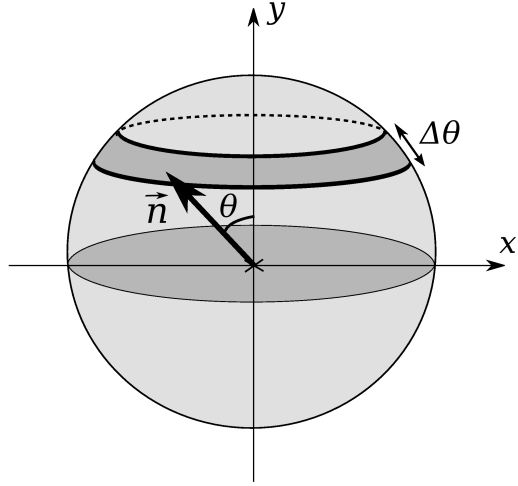}
\caption{Description of the method used for interaction orientation distribution.}
\label{fig.7}
\end{figure}

The search has been done considering the given direction angle $\theta$ as presented in Fig.\ref{fig.7}, with $\theta$ corresponding to the angle of the unit normal vector $(\vec n )$ from the axis of axisymmetry of the sample ($Y$).
As seen in section 2.2, $\sigma_{ij}^{cap}$ can be written as:
\begin{equation}
\sigma_{ij}^{cap} = \frac{1}{V} \sum_{p=1}^N \sum_{q=1}^N F_{cap}^{q,p} l^{q,p} n_i^{p,q} n_j^{p,q}
\end{equation}
which points out the possible induced anisotropy of the capillary tensor by means of both liquid bridge and force intensity distributions.
If we now introduce $P_{meniscus}(\vec{n})$ as the menisci orientation distribution inside the sample (in fact, the number of menisci along the direction $\vec{n}$) defined by $\int_{V} P_{meniscus}(\vec{n}) dV =1 $, Eq.(22) becomes:
\begin{equation}
\sigma_{ij}^{cap} = \frac{1}{V} \int_{V} <F_{cap}.l>_{\vec{n}} P_{meniscus}(\vec{n}) \vec{n} \otimes \vec{n} dV
\end{equation}
where $<F_{cap}.l>_{\vec{n}}$ is the mean value of $\vec{F_{cap}}.\vec{l} $ along the direction $\vec{n}$. It is therefore possible to compute separately the geometric distribution ($P_{meniscus}(\vec{n})$) and the static distribution of the ($<F_{cap}.l>_{\vec{n}}$) quantity which involves the mean force intensity, for every direction characterized by $\vec{n}$.\\

\begin{figure}
\centering
\includegraphics[width=130mm]{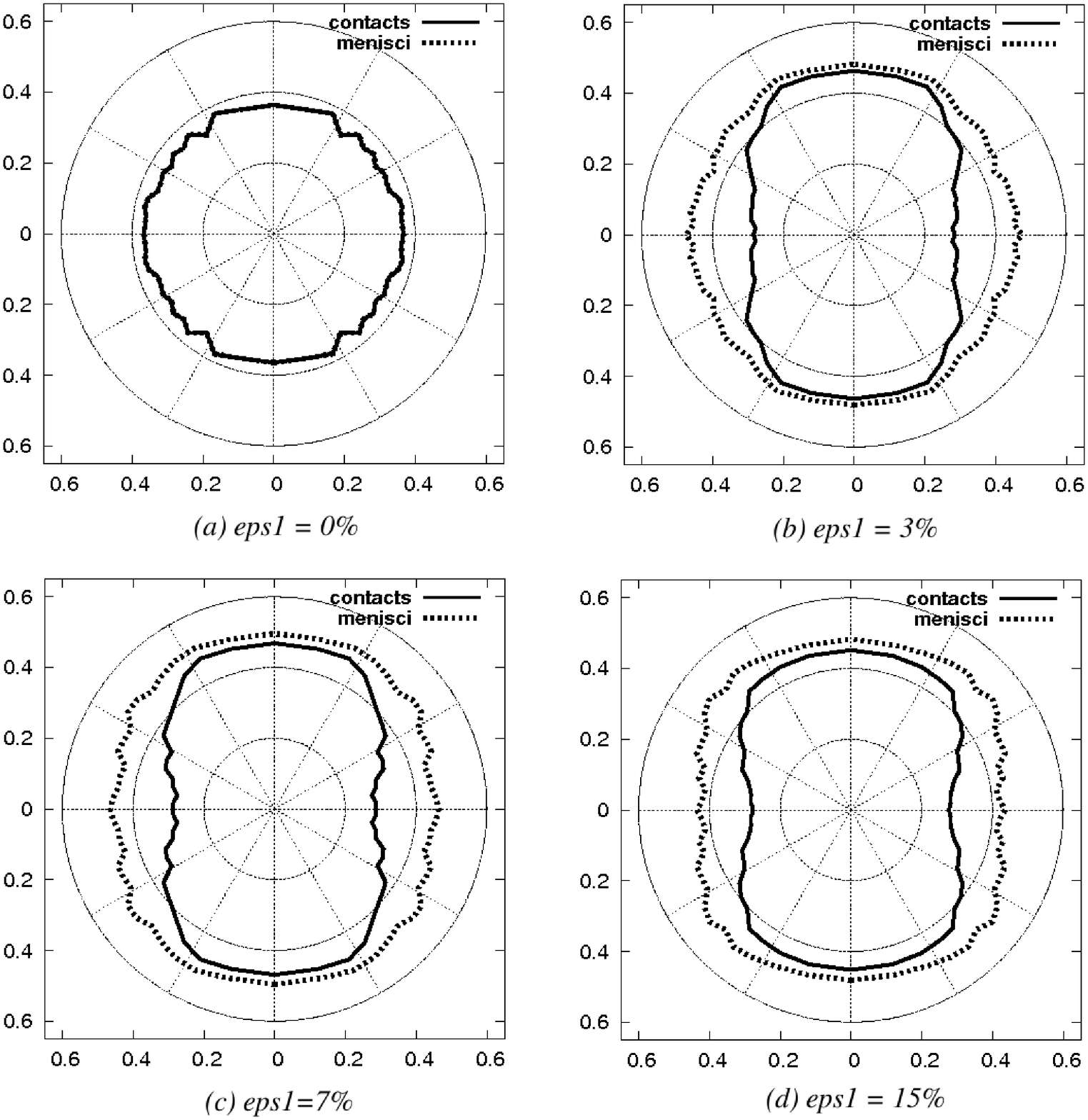}
\caption{Contacts and menisci orientation distribution $P(\vec{n})$ for different deformation levels}
\label{fig8}
\end{figure}

\begin{figure}
\centering
\includegraphics[width=130mm]{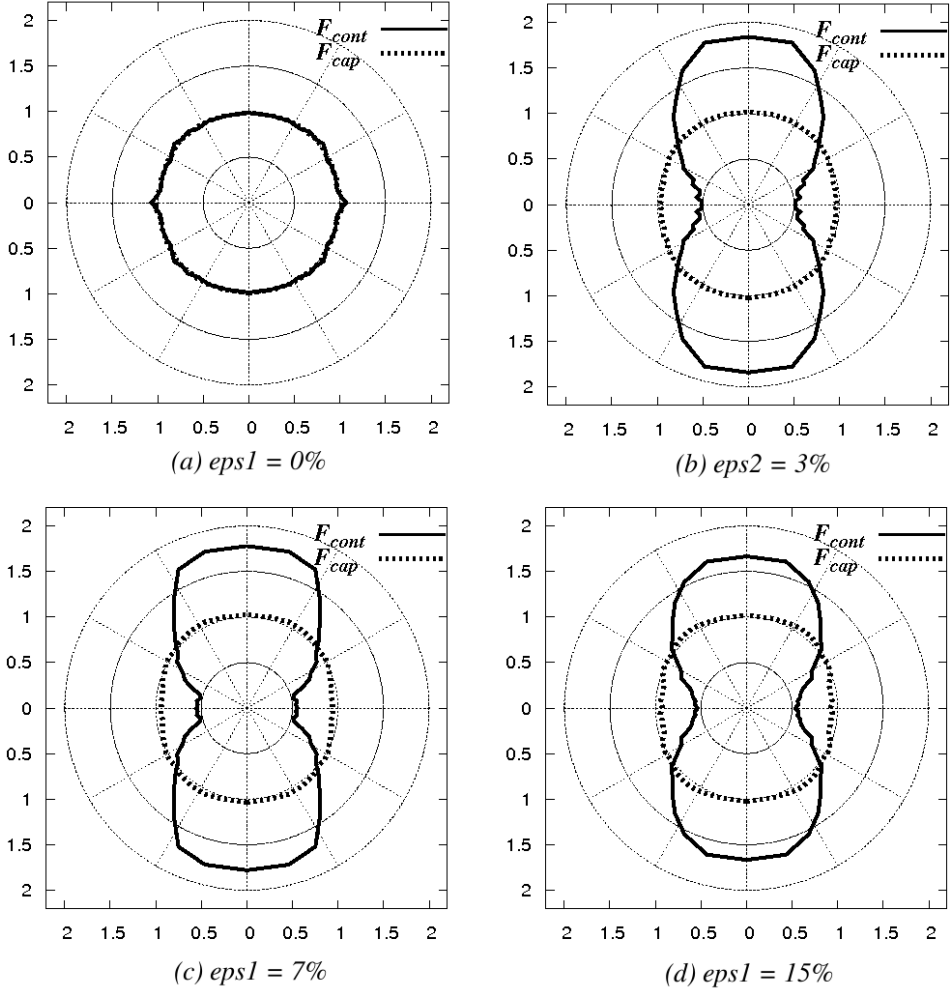}
\caption{Contacts and capillary forces distribution orientations ($<F.l>_{\vec{n}}$) for different deformation levels}
\label{fig9}
\end{figure}

$P(\vec{n})$ (Fig.\ref{fig8}) and $<F.l>_{\vec{n}}$ (fig.\ref{fig9}) distributions with $\theta$ are plotted for several deformation levels for both contacts and menisci contributions. $<F_{cont}.l>_{\vec{n}}$ and $<F_{cap}.l>_{\vec{n}}$ are simply normalized by their mean value to be qualitatively compared. The search has been done for different deformation levels on the sample confined under $15\,kPa$ and subjected to a constant capillary pressure of $10\,kPa$. Different snapshots have been taken, starting from the assumed isotropic initial state, until a $15\,\%$ axial strain level where the deformation regime appears almost permanent.
First, as menisci are added at contacts to simulate capillary condensation, distributions of both contact and capillary terms are identical in the initial state. The structural isotropy of the sample clearly appears with a uniform distribution of $P(\vec{n})$ and $<f.l>_{\vec{n}}$ for contacts and menisci along all the directions (Figs.\ref{fig8}(a) and \ref{fig9}(a)), confirming here the accuracy of the generation process.\\

The evolutions of both contact and liquid bridge distributions during the loading directly results from the deformations of the assembly. As the sample reacts just like a dense granular material (Fig.\ref{fig6}(a)), the initial contractancy gives rise to a brief increase in the coordination number (Fig.\ref{fig6}(b)), leading to the development of new liquid bridges. The corresponding growth of both axial and lateral capillary stress components can be viewed on Fig.\ref{fig5}. As a consequence of the persistance of menisci for low interparticle distances, this combined augmentation persists until $\epsilon_1 = 3\,\%$ even if $K$ strongly drops before rising up again. The small difference between the axial and the lateral components is simply due to the deviatoric loading which produces more contacts, and consequently more menisci, in the active loading direction ($Y$) than in the passive stress-controlled one.\\

After $\epsilon_1 = 3\,\%$, the lateral capillary tensor component clearly starts to reduce. As pointed out in Fig.\ref{fig8}(b), this results from the lateral spreading of the particles produced by the dilatancy of the assembly. Even though menisci distribution $P_{meniscus}(\vec{n})$ seems not to follow the induced fabric anisotropy because of the remaining of the liquid bridges, lateral capillary forces tend to diminish due to increasing interparticular distances (Fig.\ref{fig2}). On the other hand, the axial component of the capillary stress tensor regularly rises until the initiation of a permanent regime in the deformation process (near $\epsilon_1 = 7\,\%$) where $K$ continues on going quasi-constant. From this state, as the number of contacts stabilizes while dilatancy persists, $<F_{cap}.l>_{\vec{n}}$ cannot endure any further increase and the axial component of the capillary tensor starts to diminish because of a higher spreading of the grains.\\

Describing the fabric induced anisotropy, $P_{cont}(\vec{n})$ distribution from $3$ to $15\,\%$, seems to be quite constant. However, it can be noted that $<F_{cont}.l>_{\vec{n}}$ contribution appears to reach rapidly a maximum anisotropic state before slightly reducing to the final one. This maximum anisotropic strength state fairly corresponds to the peak shear strength of the sample ($\epsilon_1 = 3\,\%$) where vertical contact force chains are subjected to maximum loading before breaking off. Concerning the final distributions, they are well representative of the so-called critical state, where a stabilization of the stresses occurs with no new significative change in the evolution of the anisotropy. In a noteworthy way, menisci seem not to follow the same evolution. Effectively, as liquid bridges can exist in a certain range of increasing intergranular distances, their distribution in the media is not driven in the same way by the fabric induced anisotropy and tends to stay close to the initial state, particularly for the range of small deformations. Nevertheless, for large deformations,due to the sample dilatancy, a small induced anisotropy arises from the disappearance of liquid bridges in the lateral directions. It is evident that the history of the material is fundamentally essential when dealing with water distribution.\\

To sum up, the analysis reveals a slight induced anisotropy of the capillary stress tensor as a function of the medium fabric. The pore fluid in unsaturated soil has its own fabric that may be readily altered with changes in the granular fabric and is also strongly dependent on the water distribution inside the media. The global approximation which characterizes water effects in unsaturated materials by an equivalent pore pressure is, therefore, unable in essence to point out this intrinsically anisotropic microstructural force contribution. However, depending on the hydric history, the evolving anisotropy of the pore water distribution can validate the assumption by counterbalancing the induced fabric anisotropy in the material.\\

Since DEM analyses result from direct simulations of a granular assembly, the purpose of the next section is to compare the results with those of a microstructural model where the behaviour of the material is obtained through a micromechanically based constitutive relation.

\section{COMPARISON WITH A MICRO-MECHANICAL MODEL}
In this section, we first present the microstructural model used to compare with DEM simulations. This is a stress-strain model (\cite{Cambou1989, NematNasser2000, NicotDarve2005}) proposed by Chang and Hicher \cite{Chang2005} which considers inter-particle forces and displacements. Its capability has been recently extended to unsaturated states by incorporating the influence of capillary forces at the micro level \cite{HicherChang2006}. By comparing the predicted triaxial loading results  obtained by the two approaches for the granular assembly of section 3, we confirm the stress conceptions introduced previously, focusing on the capillary stress tensor as defined in section 2.2.

\subsection{Stress-Strain Model}
In this model, we envision a granular material as a collection of particles. The deformation of a representative volume of this material is generated by the mobilisation of contact particles in all orientations. Thus, the stress-strain relationship can be derived as an average of the mobilisation behaviour of local contact planes in all orientations.
The forces and movements at the contact planes of all orientations are suitably superimposed to obtain the macroscopic stress strain tensors using the static homogenization presented in Section 2.2.

\subsubsection{Inter-particle behaviour}
\begin{itemize}
\item[-]\textit{Contact forces and capillary forces}
\end{itemize}

For dry samples, contact forces are directly determined from the external stresses $\sigma$ applied on the granular assembly (Eq.(2)). In the case of wetted samples, different stages of saturation can be identified. The fully saturated regime corresponds to a two-phase material with water filling completely the voids between grains. The water pressure $u_w$ can either be positive or negative (suction) but in both cases the effective stress concept \cite{Terzaghi1925} can be applied and the contact forces determined by considering the effective stresses $\sigma'$ as the external stresses (\cite{DeBuhan1996, Hicher1998}):
\begin{equation}
\sigma' = \sigma - u_w I
\end{equation}
As seen previously, in the case of partially saturated samples in the pendular regime, the liquid phase is distributed in menisci located between close grains. As a consequence, capillary forces are applied on the grains and are added to the contact forces defined above. The attractive capillary force between two grains connected by a water bridge is a decreasing function of the distance between the grains until the bridge fails (Fig.\ref{fig2}). This function depends on the volume of liquid found between the grains. Different mathematical expressions have been proposed for these capillary forces, $F_{cap}$. Eq.(18) presents the expression used in DEM. $F_{cap}$ depends on the capillary pressure defined as the pressure jump across the liquid-air interface, on the liquid-air interface surface tension, as well as on the geometry of the menisci governed by the solid-liquid contact angle and the filling angle. One can see that $F_{cap}$ depends on the geometry of the liquid bridge which is function of the amount of the pore water and of the distance between two neighboring grains. The use of Eq.(18) for determining the amplitude of the capillary forces is not a straightforward one and therefore, in the micro-mechanical model, a simplified approach is to consider an empirical relation between $F_{max}$ and the degree of saturation Sr:
\begin{equation}
F^n_{cap} = F_{max} e^{-c (\frac{D}{R})}
\end{equation}
where $F^n_{cap}$ is the capillary force between two neighbouring grains, not necessarily in contact, $F_{max}$ the value of $F^n_{cap}$ for two grains in contact and $R$ the mean grain radius. $D$ represents the distance between two grains and is equal to $l-2R$, $l$ being the branch length given as a distribution function of the grain size and the void ratio and $c$ is a material parameter, dependent on the grain morphology and on the water content:
\begin{equation}
\begin{array}{ll}
F_{max} = F_0 \frac{S_r}{S_0} \quad \textrm{for} \quad 0 < S_r < S_0 \\
F_{max} = F_0 \frac{S_0 (1-S_r)}{S_r (1-S_0)} \quad \textrm{for} \quad S_0 < S_r < 0
\end{array}
\end{equation}
where $F_0$ and $S_0$ are material parameters. $F_0$ depends on the grain size distribution, $S_0$ represents the degree of saturation at which any further drying of the specimen will cause substantial breaking of the menisci in the pendular domain. $S_0$ depends on the nature of the granular material.\\

Since the menisci are not necessarily all formed in the funicular regime, Eq.(25) may not be applicable for high degrees of saturation. However, in this first approach, we decided to extend it to the whole range of saturation, considering that the amplitudes of capillary forces were small for degrees of saturation higher that 80\% and could thus be approached with sufficient accuracy by using the same equation.

\begin{itemize}
\item[-]\textit{Elastic relationship}
\end{itemize}

The contact stiffness of a contact plane includes normal stiffness, $k_n^\alpha$ , and shear stiffness, $k_t^\alpha$. The elastic stiffness tensor is defined by
\begin{equation}
F_i^\alpha = k_{ij}^{\alpha e} \delta_{j}^{\alpha e}
\end{equation}
which can be related to the contact normal and shear stiffness by
\begin{equation}
k_{ij}^{\alpha e} = k_n^{\alpha} n_i^\alpha n_j^\alpha + k_t^{\alpha} (s_i^\alpha s_j^\alpha + t_i^\alpha t_j^\alpha)
\end{equation}
The value of the stiffness for two elastic spheres can be estimated from Hertz-Mindlin's fomulation. For sand grains, a revised form was adopted \cite{Chang1989}, given by
\begin{equation}
k_n = k_{n0} (\frac{F_n}{G_g l^2})^n \quad k_t = k_{t0} (\frac{F_n}{G_g l^2})^n
\end{equation}
where $G_g$ is the elastic modulus for the grains, $F_n$ is the contact force in normal direction, $l$ is the branch length between the two particles, $k_{n0}$, $k_{t0}$ and $n$ are material constants.

\begin{itemize}
\item[-]\textit{Plastic relationship}
\end{itemize}

Plastic sliding often occurs along the tangential direction of the contact plane with an upward or downward movement, thus shear dilation/contraction takes place. 
The dilatancy effect can be described by
\begin{equation}
\frac{d \delta_n^p}{d \Delta^p} = \frac{T}{F_n} - tan \phi_0
\end{equation}
where $\phi_0$ is a material constant which, in most cases, can be considered equal to the internal friction angle $\phi_{\mu}$. This equation can be derived by equating the dissipation work due to plastic movements and friction in the same orientation. Note that the shear force $T$ and the rate of plastic sliding $d \Delta^p$ can be defined as $T = \sqrt{F_s^2 + F_t^2} \quad and \quad d \Delta^p =\sqrt{(d \delta_s^p)^2 + (d \delta_t^p)^2}$. The yield function is assumed to be of the Mohr-Coulomb type,
\begin{equation}
F(F_i, \kappa) = T - F_n \kappa (\Delta^p) = 0
\end{equation}
where $\kappa(\Delta^p)$ is an isotropic hardening/softening parameter defined as:
\begin{equation}
\kappa = \frac{k_{p0} tan (\phi_p) \Delta^p }{\vert F_n \vert tan (\phi_p) + k_{p0} \Delta^p}
\end{equation}
The hardening function is defined by a hyperbolic curve in $\kappa - \Delta^p$ plane, which involves two material constants: $\phi_p$ and $\kappa_{p0}$. 
On the yield surface, under a loading condition, the shear plastic flow is determined by a normality rule applied to the yield function. However, the plastic flow in the direction normal to the contact plane is governed by the stress-dilatancy equation in Eq.(32). Thus, the flow rule is non-associated.

\begin{itemize}
\item[-]\textit{Interlocking influence}
\end{itemize}

The internal friction angle $\phi_\mu$ is a constant for the material. However, the peak friction angle, $\phi_p$, on a contact plane is dependent on the degree of interlocking by neighboring particles, which can be related to the state of the packing void ratio $e$ by:
\begin{equation}
tan(\phi_p) = (\frac{e_c}{e})^m tan(\phi_\mu)
\end{equation}
where $m$ is a material constant \cite{Biarez1994}. The state of packing is itself related to the void ratio at critical state $e_c$. The critical void ratio $e_c$ is a function of the mean stress. The relationship has traditionally been written as:
\begin{equation}
e_c = \Gamma - \lambda log(p') \quad or \quad e_c = e_{ref} - \lambda log(\frac{p'}{p_{ref}})
\end{equation}
where $\Gamma$ and $\lambda$ are two material constants and $p'$ is the mean stress of the packing, and ($e_{ref}$, $p_{ref}$) is a reference point on the critical state line.\\
For dense packing, the peak frictional angle $\phi_p$ is greater than $\phi_\mu$. When the packing structure dilates, the degree of interlocking and the peak frictional angle are reduced, which results in a strain-softening phenomenon.

\begin{itemize}
\item[-]\textit{Elasto-plastic relationship}
\end{itemize}

With the elements discussed above, the final incremental stress-strain relations of the material can be derived that includes both elastic and plastic behaviour, given by $\dot F_i^\alpha = k_{ij}^{\alpha p} \dot \delta_j^{\alpha}$. Detailed expression of the elasto-plastic stiffness tensor is given in \cite{Chang2005}.

\subsubsection{Stress-strain relationship}

\begin{itemize}
\item[-]\textit{Macro micro relationship}
\end{itemize}

The stress-strain relationship for an assembly can be determined from integrating the behaviour of inter-particle contacts in all orientations. During the integration process, a relationship is required to link the macro and micro variables. Using the static hypotheses proposed by Liao et. al \cite{Liao1997}, we obtain the relation between the macro strain and inter-particle displacement (finite strain condition not being considered here):
\begin{equation}
u_{i,j} = A_{ik}^{-1} \sum_{\alpha = 1}^N \delta_j^\alpha l_k^\alpha
\end{equation}
where $\delta_j^\alpha$ is the relative displacement between two contact particles and the branch vector $l_k$ is the vector joining the centers of two contacting particles.

Using both the principle of energy balance and Eq.(36), the mean force on the contact plane of each orientation is
\begin{equation}
F_i^{\alpha} = \sigma_{ij} A_{jk}^{-1} l_k^\alpha V
\end{equation}
The stress increment $\sigma_{ij}$ induced by the loading can then be obtained through the contact forces and branch vectors for contacts in all orientations \cite{Christofferson1981, Rothenburg1981}. Since $l_k^\alpha$ represents the mean branch vector for the $\alpha^{th}$ orientation including both contact and non-contact particles, the value of $F_i^\alpha$ in Eq.(37) represents the mean of contact forces in the $\alpha^{th}$ orientation.
\begin{equation}
\sigma_{ij} = \frac{1}{V} \sum_{\alpha =1}^N F_i^{\alpha} l_j^\alpha 
\end{equation}
When the defined contact force is applied in Eq.(37), Eq.(38) is unconditionally satisfied.\\

Using the definition of Eq.(38), the stress induced by capillary forces can be computed and is termed as capillary stress, given by
\begin{equation}
\sigma_{ij}^{cap} = \frac{1}{V} \sum_{\alpha =1}^N F_{cap,i}^{\alpha} l_j^\alpha
\end{equation}
As mentionned in section 2.2, it is noted that this term is not analogous to the usual concept of capillary pressure or suction which represents the negative pore water pressure inside the unsaturated material. In agreement with the results obtained through DEM simulations, the capillary stress depends on the geometry of the pores and is a tensor rather than a scalar. Only for an isotropic distribution of the branch lengths $l^\alpha$, this water associated stress can be reduced to an isotropic tensor. This is the case for an initially isotropic structure during isotropic loading, but during deviatoric loading, an induced anisotropy is created and the capillary tensor is no longer isotropic.\\

At the equilibrium state, the effective intergranular forces will therefore be the difference between the repulsive forces due to the external stresses (Eq.(38)) and the attractive capillary forces (Eq.(39)). In a similar way to Eq.(7), we can thus define a generalized intergranular stress tensor $\sigma^*$, defined by:
\begin{equation}
\sigma^* = \sigma - \sigma^{cap}
\end{equation}
Assuming that $\sigma^*$ can stand as an appropriate definition of the effective stress, this equation could represent a generalisation of Eq.(1) proposed by Bishop, in which the capillary stress is reduced to an isotropic tensor. Dangla et al. \cite{Dangla1998} demonstrated the validity of the effective stress approach in elasticity by means of an energy approach. They obtained an expression similar to Eq.(40) but with an additional term corresponding to the work of the interfaces. As pointed out before, capillary forces in our models, and  consequently the capillary stresses, depend on the negative pore water pressure, or suction, and on the water-air interface surface tension. A similar approach can be found in the work of Fleureau and al. \cite{Fleureau2003} who obtained an explicit expression of the capillary stress as a function of the suction for regular arrangements of spherical grains by neglecting the surface tension. The definition of the capillary stress tensor in Eq.(39) can therefore be considered as an extension of the results obtained from these previous studies to cases of non isotropic granular assemblies.

\subsubsection{Summary of parameters}
One can summarize the material parameters as:
\begin{itemize}
\item[-] Normalized contact number per unit volume: $\frac{Nl^3}{V}$.
\item[-] mean particle size, $2R$.
\item[-] Inter-particle elastic constants: $k_{n0}$, $k_{t0}$ and $n$.
\item[-] Inter-particle friction angle: $\phi_\mu$ and $m$.
\item[-] Inter-particle hardening rule: $k_{p0}$ and $\phi_0$.
\item[-] Critical state for packing: $\lambda$ and $\Gamma$ or $e_{ref}$ and $p_{ref}$.
\item[-] Capillary force equation: $f_0$, $S_0$ and $c$.
\end{itemize}
Other than critical state parameters, all parameters are inter-particles. Standard values for $k_{p0}$ and $\phi_0$ are the following: $k_{p0} = k_{n0}$ and $\phi_0 = \phi_\mu$ and a typical ratio $\frac{k_{t0}}{k_{n0}} = 0.4$ can generally be assumed \cite{HicherChang2006}. Therefore, for dry or saturated samples, only six parameters have to come from experimental results and these can all be determined from the stress-strain curves obtained from drained or undrained compression triaxial tests. For unsaturated states, three more parameters need to be determined, using specific triaxial tests on partially saturated samples.

\subsection{Numerical simulations}
Several simulations of triaxial loading have been performed in order to compare DEM and Micromechanical Model results. These are based on computations on a Representative Volume Element of about 10 000 spherical elements.
\subsubsection{Dry samples}
\begin{figure}
\centering
\centering
\includegraphics[width=100mm]{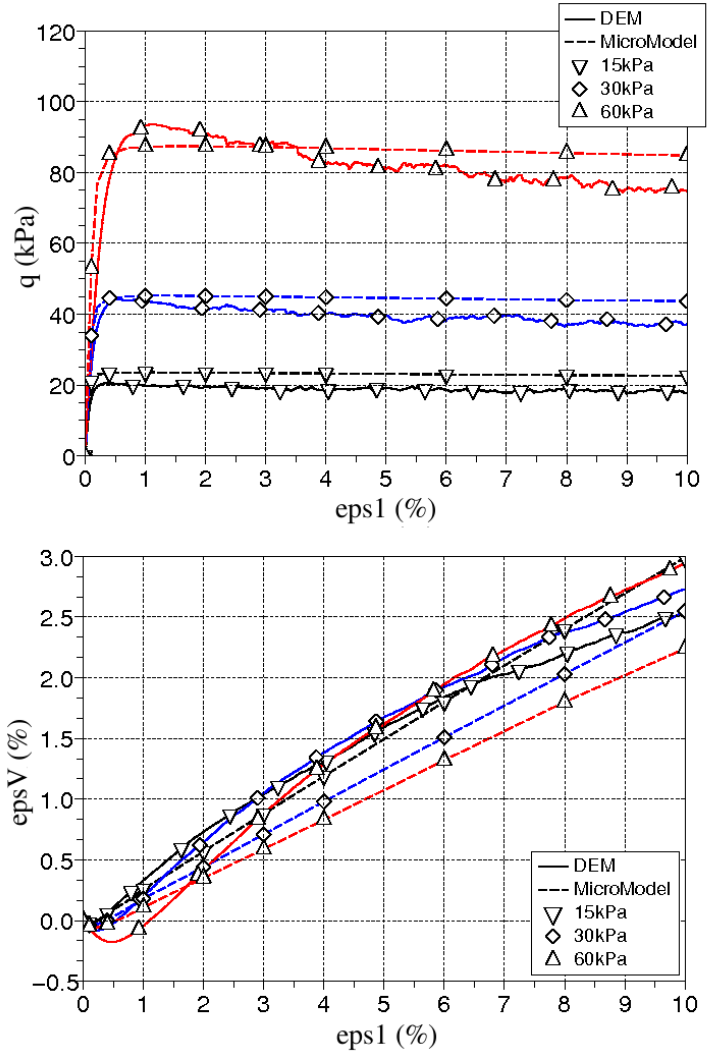}
\caption{DEM and MicroMechanical simulations of triaxial compression tests on a similar dry granular assembly.}
\label{fig11}
\end{figure}

The model parameters were determined from the following procedure. The granular assembly defined in the discrete element model is made of spherical particles with the grain size distribution presented in Fig.\ref{fig3}. The particle size $2R$ was selected equal to $d_{50} = 0.045\,mm$. The elastic parameters could not be directly derived from the inter-particle behaviour used in the DEM simulations, which consider a linear contact stiffness. From previous studies on glass beads assemblies \cite{Hicher1996, HicherChang2006}, typical values were adopted in this study. The plastic parameters were determined from the numerical results obtained by DEM. Table II summarizes the set of parameters used for modelling the dry sample behaviour.
\begin{table}[h]
\begin{center}
\begin{tabular}[c]{c c c c c c}
\hline
$k_{n0}$ (N/m)	& $\frac{k_{t0}}{k_{n0}}$	& $n$	&$\phi_\mu$ (deg.)	&$\lambda$	&$m$\\
\hline
300		& 0.5				& 30	&20			&0.05		&0.5 \\
\hline
\end{tabular}
\caption{Microstructural Model parameters for the glass beads assembly}
\end{center}
\end{table}\\
Fig.\ref{fig11} presents the numerical simulations of three triaxial tests performed at three different confining pressures of $15$, $30$ and $60\,kPa$ for the simulated dry assembly having an initial void ratio equal to $0.38$. One can see that the results obtained with the Microstructural Model compared well to DEM ones.

\subsubsection{Wet samples}
DEM simulations were performed on unsaturated assemblies with an initial saturation degree of about $20\,\%$, corresponding to a suction value equal to $10\,kPa$. In order to determine the corresponding capillary forces, Eq.(25) includes two material parameters $c$ and $d$ which control the evolution of the water induced forces with the distance between two particles. According to experimental results presented in different studies \cite{Lian1993, Soulie2006}, we decided to take a value of  $c = 4$ and of $d = 0.05\,mm$. These values give a standard evolution for the capillary forces, function of the distance between particles, as well as an initial isotropic distribution of these forces if the material structure is isotropic. The evolution of the capillary forces with the degree of saturation requires two more parameters: $S_0$ and $f_0$. According to previous studies \cite{HicherChang2006}, we selected a value of $S_0 = 1\,\%$ and determined the value of $f_0 = 0.045\,N$. In order to obtain an initial value of the capillary stress in accordance with the one computed by DEM. We then performed numerical simulations of triaxial tests on wet samples for different confining pressures. Contrary to DEM suction-controlled simulations, the Microstructural Model tests are water content controlled. In order to compare those two approaches, samples were initially wetted at a common degree of saturation of about $20\,\%$.
\begin{figure}
\centering
\includegraphics[width=90mm]{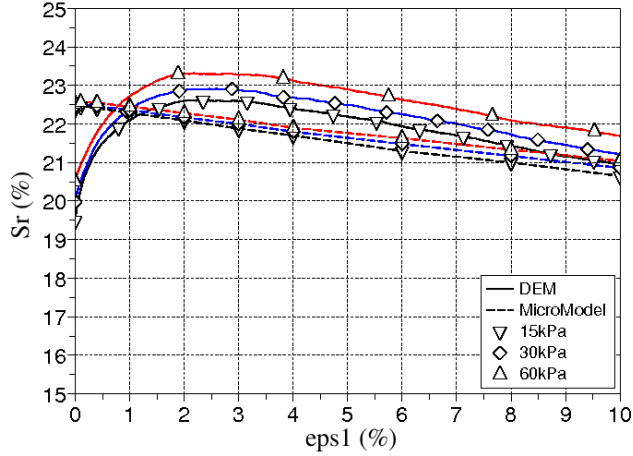}
\caption{Evolution of the saturation degree during DEM and Micromechanical simulated triaxial tests.}
\label{fig12}
\end{figure}
One should notice that, even if the test conditions are not strictly identical, the changes in the degree of saturation during loading obtained for both tests are sufficiently similar for us (Fig.\ref{fig12}) to compare the results obtained by the two approaches.
\begin{figure}
\centering
\includegraphics[width=95mm]{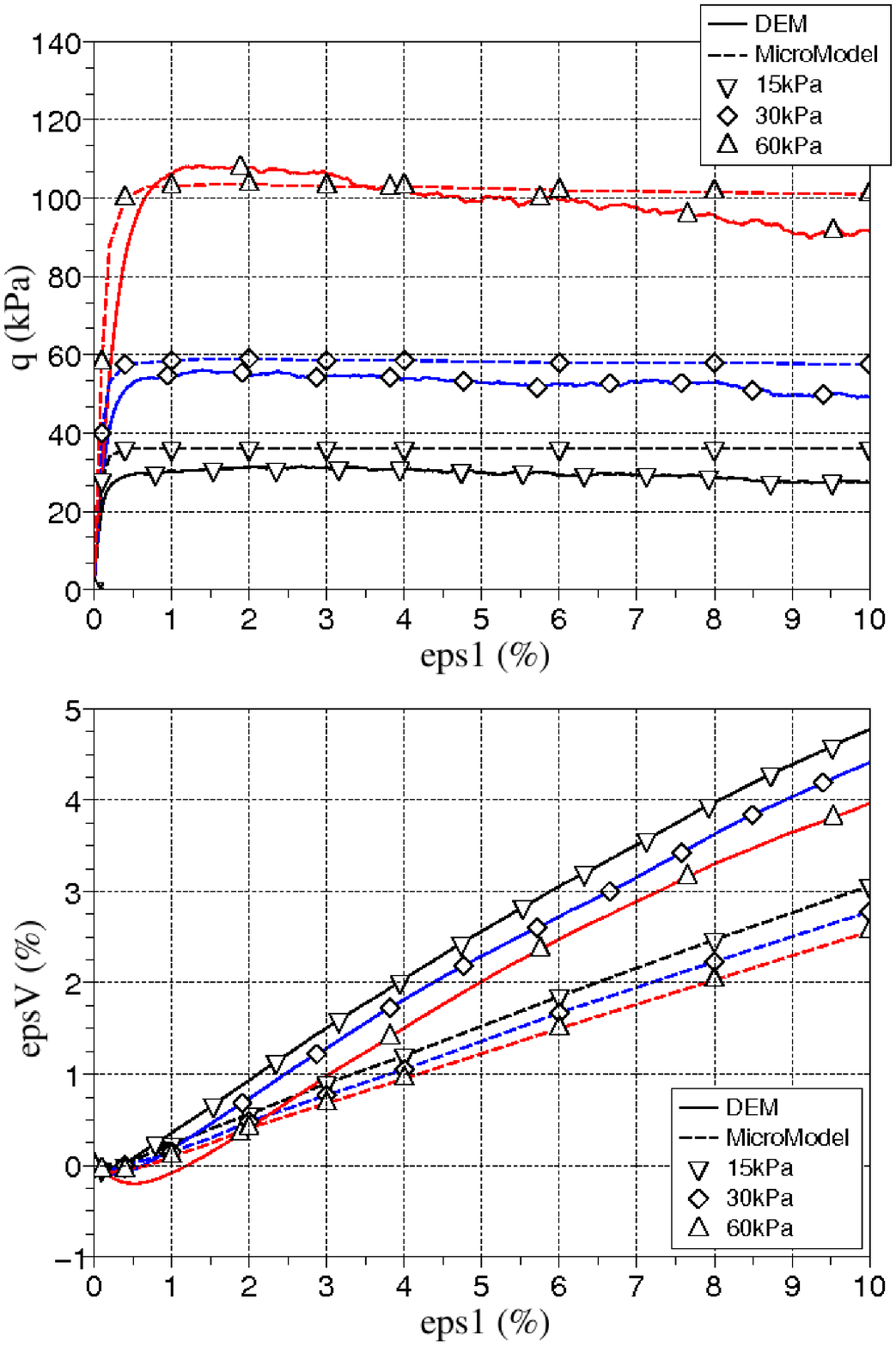}
\caption{DEM and Micromechanical simulations of triaxial compression tests on a similar wet granular assembly.}
\label{fig13}
\end{figure}\\

As presented in Fig.\ref{fig13}, the two models give quite similar results. One can see that a material strength increase is obtained for unsaturated samples compared to dry ones at the same confining pressure. The volume changes during triaxial loading create a small change in the degree of saturation (Fig.\ref{fig12}). As a consequence, the capillary forces evolve, according to Eq.(5). 
\begin{figure}
\centering
\includegraphics[width=100mm]{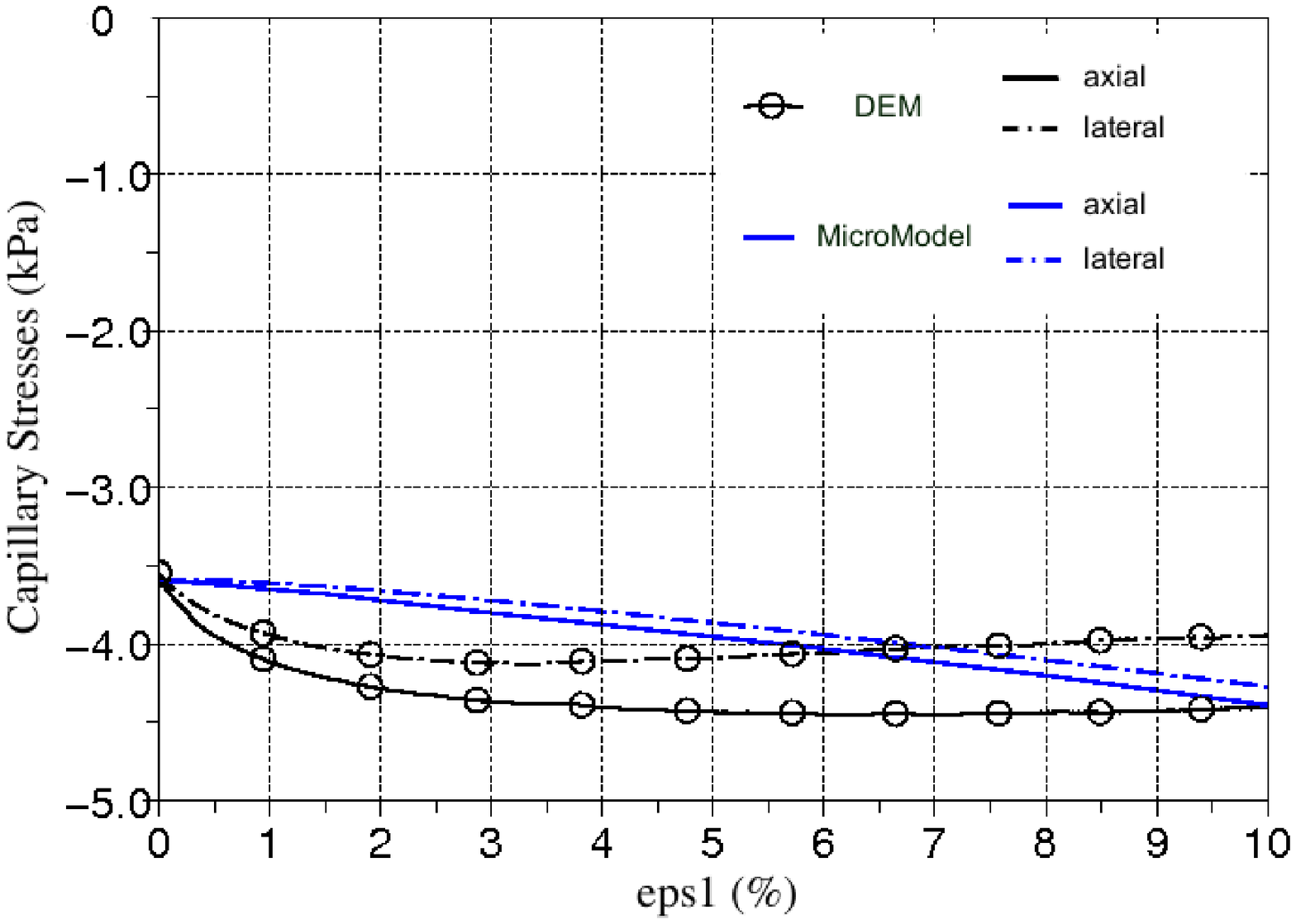}
\caption{Evolution of the principal capillary stress tensor components during a triaxial compression test on a wet granular assembly.}
\label{fig14}
\end{figure}
Fig.\ref{fig14} shows the evolution of the principal components of the capillary stress tensor during constant water content triaxial tests. The initial state corresponds to an isotropic capillary stress tensor with a mean stress equal to $3.6\,kPa$ as obtained in DEM. During loading, a structural anisotropy is created due to the evolution of the fabric tensor defined in Eq.(38). Therefore, the principal components of the capillary stress tensor evolve differently. In the studied cases, this difference remains small and corresponds at the end of the test to a relative difference less than $10\,\%$. This small difference can be explained by two causes. The first one is the small amount of induced anisotropy obtained by the evolution of the fabric tensor. This evolution is due to the change in the branch length $l_i^\alpha$ for each $\alpha$ direction which, in this version of the model, is only due to elastic deformations of the grains in contact. Since all our numerical testing were performed at small confining stresses, the amount of elastic deformation remained quite small. The second reason is linked to the fact that the capillary bridges can exist for non-touching neighboring grains. This has been taken into account in calculating the mean capillary force and also in determining the capillary stress tensor (Eqs.(27) and (41)). This result is in agreement with the distribution of the contacts and menisci distribution computed by DEM (Figs.\ref{fig8} and \ref{fig9}).\\

Regarding the constitutive behaviour at contacts between solid particles, the results provided by the two methods are in rather good agreement concerning the influence of capillary forces at a macroscopic level. The increase in the shear strength classically observed for partially saturated materials is clearly encountered starting from microscopical considerations, and the slight induced anisotropy of the capillary stress tensor is observed, confirming that suction effects in unsaturated materials cannot be precisely accounted for by an equivalent pore pressure assumption.

\section{CONCLUSION}
Starting from local capillary forces, a stress variable, denoted as the capillary stress tensor and intrinsically associated to water effects, has been defined through homogenisation techniques. Triaxial compression test simulations from two fundamentally different micromechanical models were performed on a granular assembly under several confining pressures for dry and partially saturated conditions. Both models reproduce in quite good agreement the main features of unsaturated granular materials, in particular the increase of the shear strength due to capillary effects.\\
The results also suggest that, in partially saturated materials within the pendular regime, the effects of pore fluid are adequately represented by a discrete distribution of forces rather than by an averaged pressure in the fluid phase. Effectively, as a representative quantity of the pore fluid distribution inside unsaturated materials, this suction associated stress tensor reveals that pore fluid has its own fabric which is inherently anisotropic and strongly dependent on the combined loading and hydric history of the material. Even if the induced anisotropy of the capillary stress tensor appears slight in this study, it is obvious that this tensorial nature of water in unsaturated material implies suction to produce shear effects on the solid phase. This suction induced shear effect consequently makes it difficult to associate an isotropic quantity to water as expressed in the Bishop's effective stress. Pore pressure is no longer an isotropic stress in unsaturated soil, and therefore, cannot be considered as an equivalent continuum medium. The analysis finally confirms that suction is a pore-scale concept, and that stress definitions for unsaturated soils should also include microscopic interparticle stresses as the ones resulting from capillary forces.\\
The multi-scale approach presented here appears to be a pertinent complementary tool for the study of unsaturated soil mechanics. More precisely, discrete methods should convey new insights into the discussion about the controversial concept of generalized effective stress by relating basic physical aspects to classical phenomenological views.


\end{document}